\documentclass[11pt,a4paper]{article}

\usepackage[french, english]{babel}
\usepackage{tikz}
\usepackage{helvet}
\usepackage{multirow}
\usepackage{amsmath}

\newcommand{\R}{\mathbb{R}}

\usepackage{amsfonts, amssymb}
\usepackage[figuresright]{rotating}
\usepackage{bbm}

\newcommand{\RN}[1]{
  \textup{\uppercase\expandafter{\romannumeral#1}}
}

\def\bSig\mathbf{\Sigma}

\usepackage{kbordermatrix}
\renewcommand{\kbldelim}{(}
\renewcommand{\kbrdelim}{)}
\renewcommand{\kbrowstyle}{\small} 
\renewcommand{\kbcolstyle}{\small}

\usepackage{subfig}
\usepackage{float}

\usepackage{xcolor}
\usepackage{bigints}

\usepackage{url}

\usepackage{graphicx}
\usepackage[myheadings]{fullpage}
\usepackage{epsfig,epsf}
\usepackage{amssymb}
\usepackage[nolists]{endfloat}
\usepackage{setspace}
\usepackage{amsmath}

\usepackage{txfonts}

\usepackage{apacite}
\usepackage[sort&compress,authoryear]{natbib}

\usepackage{prodint}
\graphicspath{{anc/}}

\begin{document}

\title{Individual dynamic predictions using landmarking and joint modelling: validation of estimators and robustness assessment}

\author{Lo\"ic Ferrer$^{1,\dagger}$, Hein Putter$^{2}$ and C\'ecile Proust-Lima$^{1}$}

\date{June 2018}

\maketitle

\vspace{0.3cm}

\noindent 
$^{1}$ INSERM, UMR1219, Univ. Bordeaux,  ISPED, F-33076 Bordeaux, France \\
$^{2}$ Leiden University Medical Center, Leiden, the Netherlands \\
$^{\dagger}$ email: loic.ferrer@inserm.fr

\label{firstpage}

\vspace{0.9cm}
\noindent {\bf {Abstract:}}
After the diagnosis of a disease, one major objective is to predict cumulative probabilities of events such as clinical relapse or death from the individual information collected up to a prediction time, usually including biomarker repeated measurements. Several competing estimators have been proposed, mainly from two approaches: joint modelling and landmarking. These approaches differ by the information used, the model assumptions and the complexity of the computational procedures. This paper aims to review the two approaches, precisely define the derived estimators of dynamic predictions and compare their performances notably in case of misspecification. The ultimate goal is to provide key elements for the use of individual dynamic predictions in clinical practice. Prediction of two competing causes of prostate cancer progression from the history of prostate-specific antigen is used as a motivated example. We formally define the quantity to estimate and its estimators, propose techniques to assess the uncertainty around predictions and validate them. We then conduct an in-depth simulation study compare the estimators in terms of prediction error, discriminatory power, efficiency and robustness to model assumptions. We show that prediction tools should be handled with care, in particular by properly specifying models and estimators.\\


\pagebreak

\section{Introduction}
\label{s:intro}

After diagnosis and subsequent treatment of cancer, patients are typically monitored via repeated measurements of biomarkers. For example, in patients with prostate cancer treated by radiotherapy, the Prostate Specific Antigen (PSA) is measured routinely.
Precisely predicting the individualized probabilities of events such as clinical relapse for these patients from their individual information collected until a prediction time has become a central issue \citep{proust2009development, goldstein2017comparison}.
Personalized treatment strategies can indeed be proposed according to the updated individual probabilities \citep{sene2016individualized},
or the planning of the next biomarker measurement can be optimized \citep{rizopoulos2015personalized}.

Two main approaches have been proposed to compute individual dynamic predictions: the \emph{joint modelling approach} and the \emph{landmarking approach}. These techniques differ in the used information, the model assumptions and the complexity of computational procedures.

The \emph{joint modelling (JM) approach} simultaneously models the repeated measurements of a biomarker (e.g., using a linear mixed model in standard JM) and time-to-event data (e.g., using a proportional hazards model in standard JM) by linking them using a function of shared random effects \citep{tsiatis2004joint}. This approach has the advantage of taking into account the endogenous nature of biomarkers \citep{kalbfleisch2011statistical}, of only requiring one model estimation for all prediction times, and of modelling the progression of the disease as a whole, which makes it very popular.
But it is often based on simplifying assumptions (e.g., proportional hazards, number of random effects) and may be complex to estimate, so that it should be handled carefully and can remain difficult to apply in practice. 

The \emph{landmarking approach} consists of adjusting standard survival models considering only the subsample of subjects still at risk at the prediction time and the longitudinal information collected up to the prediction time \citep{van2007dynamic}. Classically, the model is a Cox model (or cause-specific proportional hazard model) with a truncation at the prediction time. As such, these models induce significantly less numerical problems. In addition a censoring is usually administered at the end of the prediction window to reduce possible bias related to the proportionality of hazards. However, as they do not fully explore the collected information during the follow-up and the correlation between the marker and the time of event, they can produce sub-efficient estimators \citep{huang2016two} and are only an approximation of the (correct) joint estimator. As explained in Suresh et al. (2017) \cite{suresh2017comparison} they do not satisfy the consistency condition introduced by Jewell and Nielsen (1993) \cite{jewell1993framework} which stipulates that the hazard function and the marker dynamics must be linked at all time points to give consistent dynamic predictions. In the presence of a longitudinal biomarker, the naive landmark approach consists in adjusting the survival model on the last observed value of the biomarker. To take into account the measurement errors of the biomarker and its collection at discrete times, the last observed value may be replaced by the predicted value at the prediction time obtained from a linear mixed model \citep{sweeting2017use, rizopoulos2017dynpred}. This two-stage approach also takes into account all the collected information of the biomarker until the prediction time for the subjects at risk. However the event probabilities must still be deduced by approximation and the model is not completely freed of the proportional hazards assumption.

In the context of competing risks, instead of assuming a cause-specific proportional hazard model, the conditional probabilities of event can be directly modelled by considering a dynamic pseudo-observations approach \citep{nicolaie2013dynamic}, which is freed from the proportionality hazards assumption. By including the predicted value of the biomarker at the prediction time as a covariate, it also takes into account the measurement errors of the biomarker and its collection in discrete time. But it requires the specification of a link function and can still provide less efficient estimators than the joint model.

Although of central interest in many recent works, there remains some vagueness in the definition of estimators of dynamic predictions and of their uncertainty. Several definitions exist in the joint modelling framework \citep{rizopoulos2011dynamic,sene2016individualized, barrett2017dynamic} and no concept of uncertainty was introduced in the landmarking framework \citep{proust2009development,sweeting2017use,rizopoulos2017dynpred}. Overall, estimators of dynamic predictions have never been formally validated in simulations.
We thus aimed to first define estimators of individual dynamic predictions, propose estimators of their uncertainty with 95\% confidence intervals for the joint modelling and the landmarking approaches and validate them in a simulation study. We then aimed to compare the predictive accuracy of the models under several scenarios to explore their robustness to misspecification. We used the prediction of competing progressions of prostate cancer from the PSA history as the motivating and illustrating example.

The rest of the paper is organized as follows.
Section~\ref{s:prediction_models} introduces the concept of individual dynamic prediction, the two modelling approaches and the derived estimators of dynamic predictions and of their uncertainty. Section~\ref{s:data} describes the motivating data and gives an illustration of computed individual dynamic predictions using landmark models or joint models. The simulation studies are carried out in Section~\ref{s:simu} for validating the proposed estimators and comparing them in terms of prediction accuracy. The paper ends with a discussion and recommendations in Section~\ref{s:discussion}.


\section{Prediction models} 
\label{s:prediction_models}

Let us consider without loss of generality the setting where subjects are at risk to experience $K$ competing events. For each subject $i$ ($i=1,...,N$), we denote $T_i$ the earliest time-to-event and $\delta_i = k$ the cause of failure ($k \in 1,\ldots,K$).
In the presence of censoring, we observe the event time $T_i^\dagger = \min(T_i, C_i)$ with $C_i$ the censoring time, and the indicator of event becomes $\Delta_i = \delta_i \, . \,\mathbbm 1\{T_i \leq C_i\}$ with $\mathbbm 1$ the indicator function. We also observe $X_i$ the (possibly time-dependent) exogenous covariates collected until the event time and $Y_i$ an endogenous longitudinal marker repeatedly measured such as $Y_i(t_{ij})$ is the observed measure at time $t_{ij}$ ($j=1,...,n_i$), with $t_{in_{i}} \leq T_i^\dagger$.

In the following, $\mathcal X_i(s)$ denotes the history of $X_i$ until time $s$, $\mathcal Y_i(s) = \{Y_i(t_{ij}): 0 \leq t_{ij} \leq s, j = 1,\ldots,n_i(s)\}$ denotes the history of the marker until $s$, and the model formulations assume a Gaussian distribution for the marker.

\subsection{Definition of individual dynamic prediction}

We are interested in the individual probability of experiencing event of cause $k$ between times $s$ and $s + w$ for a new subject $\star$ conditional to the history $\mathcal X_\star(s)$ and  $\mathcal Y_\star(s)$. Time $s$ is called the landmark time (or prediction time) and $w$ the horizon of prediction. This probability is defined as
\begin{equation}
\pi_\star^k(s,w) = \Pr(s < T_\star \leq s + w, \delta_\star = k | T_\star > s, \mathcal Y_\star(s), \mathcal X_\star(s)).
\end{equation}

We focus on models that express this quantity as a function of a vector of parameters $\theta$:
\vspace{-.15cm}
\begin{equation}
\pi_\star^k(s,w ; \theta) = \Pr(s < T_\star \leq s + w, \delta_\star = k | T_\star > s, \mathcal Y_\star(s), \mathcal X_\star(s); \theta).
\label{target_P}
\end{equation}

\vspace{-.15cm}
In practice, $\theta$ is unknown and is replaced by $\widehat\theta_\mathcal{I}$, its estimate from the considered observed data in the learning sample $\mathcal{I}$. In the remainder of the manuscript, this subscript is omitted for the sake of readability, and the estimated quantity of interest is denoted $\widehat\pi_\star^k(s,w ; \widehat\theta)$.
\subsection{Joint model}
\label{s:s:jm}

\subsubsection{Model formulation}
\label{p:jm_model}

$~$ \\
The joint model considers the full collected information 
$\mathcal I = \{(T_i^\dagger,\Delta_i, \mathcal Y_i(T_i^\dagger), \mathcal X_i(T_i^\dagger)); i=1,\ldots,N \}$.
It is decomposed into two sub-models linked by a function of a shared latent structure.
The most popular joint model \citep{rizopoulos2011dynamic} links a linear mixed model for the repeated measurements of the marker and a cause-specific proportional hazards model for the specific hazard of each cause $k$ of event using a function of shared random effects:
\begin{eqnarray}
\left\{
\begin{array}{rll}
Y_i(t) &= &m_{i}(t) + \epsilon_i(t) \nonumber \\[.1cm]
       &= & X^L_i(t)^\top \beta + Z_i(t)^\top b_i + \epsilon_i(t), \nonumber \\[.2cm]
\lambda_{i}^{k}(t) &= &\lambda_{k,0}(t) \exp\left\lbrace X^{E~\top}_{k,i} \gamma_k + W_{k,i}(t|b_i;\beta)^\top \eta_k \right\rbrace,
\end{array}
\right.
\end{eqnarray}
where $t > 0$ and $\lambda_{i}^{k}(t)$ denotes the hazard function of cause $k$ at time $t$, with $k=1,\ldots,K$.
In the longitudinal sub-part, $X^L_i(t)$ and $Z_i(t)$ denote vectors of covariates (possibly time-dependent) associated respectively with the vector of fixed effects $\beta$ and the vector of random effects $b_i, b_i \sim \mathcal N_q(0,D)$. 
The error term is $\epsilon_i(t) \sim \mathcal N(0, \sigma^2)$ ; the random effects and error terms are independent.
In the survival sub-part, $\lambda_{k,0}(t)$ denotes the parametric baseline hazard of cause $k$ at time $t$. The vector of covariates $X^{E}_{k,i}$ is associated with the vector of coefficients $\gamma_k$. We do not consider here any time-dependent exogenous prognostic variable although this is not a requirement. 
The (possibly multivariate) function $W_{k,i}(t|b_i;\beta)$ denotes the function of dependence between the longitudinal process and the hazard of event of cause $k$. Examples include the unbiased current level of the marker $m_i(t)$, the unbiased current slope $\partial m_i(t) / \partial t$ or both $\left(m_i(t), \partial m_i(t) / \partial t \right)^\top$.

A joint model can be estimated in the maximum likelihood framework using the independence between the longitudinal process $\mathcal Y_i(T_i^\dagger)$ and the survival process $(T_i^\dagger, \Delta_i)$ conditionally on the random effects $b_i$. The likelihood involves integrals over the random effects and time that have to be numerically solved, usually using Gaussian quadratures \citep{rizopoulos2012fast}. Note that the number of quadrature points has to be chosen carefully to provide correct inference \citep{ferrer2016joint}.

\subsubsection{Cumulative probability estimator}
\label{p:jm_ci}
$~$\\
Once the model is estimated, the vector of parameters $\widehat\theta$ and its variance matrix $\widehat{V(\widehat\theta)}$ are obtained, with $\widehat\theta=( \widehat\beta^\top, \widehat\sigma^{2}, \widehat\theta_{\lambda_0}^\top,\widehat\gamma^\top, \widehat\eta^\top, \textrm{vec}(\widehat D)^\top )^\top$, where $\widehat\gamma = (\widehat\gamma_1,\ldots,\widehat\gamma_K)^\top$,  $\widehat\eta = (\widehat\eta_1,\ldots,\widehat\eta_K)^\top$ and $\widehat\theta_{\lambda_0} = (\widehat\lambda_{1,0},\ldots,\widehat\lambda_{K,0})^\top$ denotes the parameters for the baseline hazards. The predicted conditional cumulative probability of cause $k$ can thus be computed for a new subject $\star$ for any landmark time $s$ and horizon $w$:
\begin{align}
\label{JM_predM}
\widehat{\pi}_\star^k(s,w; \widehat{\theta}) &= \nonumber\\
 & \hspace{-1cm}\int_{\R^q} \Pr(s < T_\star \leq s + w, \delta_\star = k | T_\star > s, \mathcal X_\star(s), b_\star; \widehat\theta) \, f(b_\star|T_\star>s, \mathcal Y_\star(s), \mathcal X_\star(s); \widehat\theta) \, \mathrm db_\star.
\end{align}
The integral \eqref{JM_predM} is usually approximated by a Gaussian quadrature; we call this estimator the \emph{marginal estimator}. When approximated by a Laplace approximation, the estimator becomes the integrand computed at the modal point; we refer to this faster but less accurate alternative as the \emph{conditional estimator}.
See details on the marginal and conditional estimators in Sections 1.1 and 1.2 of the Supplementary Material.

As the model is fully parametric, the 95\% confidence interval of \eqref{JM_predM} can be obtained using a parametric bootstrap technique. The procedure is realized as follows:\\[.2cm]
\fcolorbox{gray!30}{gray!30}{\parbox{\textwidth}{Consider a large $L$; for each $l=1,\ldots,L$,\\[.2cm]
\begin{minipage}{.98\textwidth}
\begin{enumerate}
\item[1.] generate parameters from their asymptotic distribution $\widetilde\theta^{(l)} \sim \mathcal N(\widehat\theta, \widehat{V(\widehat\theta)})$;
\item[2.] compute the predicted probability $\widetilde{\pi}_\star^{k,(l)}(s,w; \widetilde{\theta}^{(l)})$ defined in \eqref{JM_predM} for parameter values $\widetilde\theta^{(l)}$ instead of $\widehat\theta$.
\end{enumerate}
Compute the 95\% confidence interval from the 2.5th and 97.5th percentiles of
\(\{ \widetilde{\pi}_\star^{k,(l)}(s,w; \widetilde{\theta}^{(l)}); l=1,\ldots,L \}\). 
\end{minipage}}}\\[.2cm]

\subsection{Landmark cause-specific proportional hazards model}
\label{s:s:lm}

In contrast with joint models, landmark models only consider subjects at risk at a given landmark time $s$ and the longitudinal information $\{\mathcal{Y}(s),\mathcal X(s)\}$ collected until $s$.
When considering PH landmark models, administrative censoring is applied at the end of the prediction window $s+w$ in order to reduce the possible bias entailed by a violation of the PH assumption.
The considered information becomes $\mathcal I = \{( T_i^\dagger(s,w), \Delta_i(s,w), \mathcal Y_i(s), \mathcal X_i(s) ); i=1,\ldots,N^\dagger(s) \}$, with $T_i^\dagger(s,w) = \min(T_i^\dagger,s+w)$, $\Delta_i(s,w) = \Delta_i \, . \, \mathbbm 1\{ s < T_i \leq s+w \}$ and $N^\dagger(s) = \sum_{i=1}^{N} \mathbbm 1\{ T_i^\dagger > s \}$. 

\subsubsection{Model formulation}
$~$\\
The landmark cause-specific (CS) proportional hazards (PH) model is defined by 
\vspace{-.1cm}
\begin{equation*}
\lambda_{i}^{k}(t) = \lambda_{k,0}(t) \exp\left\lbrace X^{E~\top}_{k,i} \gamma_k + W_{k,i}(s)^\top \eta_k \right\rbrace,
\vspace{-.1cm}
\end{equation*}
where $t>s$, $\lambda_{k,0}(.)$ is a cause-specific baseline hazard function (most often left unspecified) and $W_{k,i}(s)$ is a multivariate function that depicts the dynamics of the marker extrapolated at  time $s$.
When $\lambda_{k,0}(.)$ is left unspecified, as it is considered in the following, the model is estimated by maximizing the Cox partial likelihood for each considered pair of landmark and horizon times. Note that for the sake of clarity, we did not use a subscript $s,w$ for the model parameters although they are different for each $(s,w)$.

To take into account the information of the marker before landmark time $s$, one can consider the last observed value only, i.e. $W_{k,i}(s) = Y_i(t_{in_i(s)})$. However, this technique, called \emph{naive landmark model} assumes that the marker is measured without error and considers neither the whole trajectory of the marker until $s$ nor the subject-specific gap between $t_{in_i(s)}$ and $s$.  A better alternative is to deduce the value of $W_{k,i}(s)$ at time $s$ from a linear mixed model estimated on the marker measurements collected until $s$ in subjects at risk at $s$. This technique is called \emph{two-stage landmark model}. For instance, by considering the same notations as in Section~\ref{p:jm_model} for the linear mixed model and the expected level of the biomarker in $s$ as the shared quantity, $W_{k,i}(s) = \widehat{Y}_i(s) = X^L_i(s)^\top \widehat \beta + Z_i(s)^\top \widehat b_i$ where $\widehat\beta$ is the vector of estimated fixed effects and $\widehat b_i = \mathbbm E ( b_i | \mathcal Y_i(s), \mathcal X_i(s) ;\widehat \theta ) = \widehat D Z_i^\top \widehat V_i^{-1} (Y_i - X^L_i \widehat \beta)$ is the vector of empirical Bayes estimates of the individual random effects, with $\widehat V_i = Z_i\widehat DZ_i^\top + \widehat \sigma^2 I_{n_i(s)}$. Here $X^L_i$ and $Z_i$ are the matrices of covariates with respectively the row vectors $X^L_i(t_{ij})^\top$ and $Z_i(t_{ij})^\top$, and the column vector $Y_i$ is with elements $Y_{ij}$, for $j= 1,\ldots, n_i(s)$. $I$ is the identity matrix. More generally, in the \emph{two-stage landmark model}, the shared quantity is $W_{k,i}(s)=\widehat W_{k,i}(s |\widehat b_i ; \widehat \beta)$.


\subsubsection{Cumulative probability estimator}
\label{p:2slm_ci}

$~$\\
With the two-stage approach, the predicted conditional cumulative probability of cause $k$ is
\vspace{-.1cm}
\begin{equation}
\widehat\pi_\star^k(s,w ; \widehat\theta) = \Pr \, ( s < T_\star \leq s+w, \delta_\star = k | T_\star > s, \mathcal X_\star(s), \widehat{b}_\star ; \widehat\theta),
\vspace{-.1cm}
\end{equation}
where $\widehat\theta=( \widehat\beta^\top, \widehat\sigma^{2}, \widehat\gamma^\top, \widehat\eta^\top, \textrm{vec}(\widehat D)^\top )^\top$ is the vector of estimated parameters (with associated estimated variance $\widehat{V(\widehat\theta)}$), and $\widehat b_\star = \mathbbm E(b_\star | \mathcal Y_\star(s), \mathcal X_\star(s); \widehat \theta)$.

To estimate valid 95\% confidence intervals, it is necessary to take into account the variability due to the parameters and baseline hazard estimates. 
The unspecified cumulative baseline hazard \(\Lambda_{k,0}(t) = \int\limits_{s}^{t} \lambda_{k,0}(u) \, \mathrm du\) is estimated using the Breslow's estimator \citep{breslow1972discussion}, \(\widehat\Lambda_{k,0}(t) = \int\limits_{s}^{t} \widehat\Pi_k^{(0)}(\widehat\theta,u)^{-1}\, \mathrm d \bar J_k(u)\) where $\widehat\Pi_k^{(0)}(\widehat\theta,u) = \dfrac{1}{N^\dagger(s)} \sum\limits_{i=1}^{N^\dagger(s)} \mathbbm 1\{ T_i^\dagger(s,w) \geq u\} \exp\left\lbrace X^{E~\top}_{k,i} \widehat\gamma_k + \widehat W_{k,i}(s | \widehat b_i ; \widehat \beta)^\top \widehat\eta_k \right\rbrace$ and $\bar J_k(u) = \dfrac{1}{N^\dagger(s)} \sum\limits_{i=1}^{N^\dagger(s)} \mathbbm 1\{ T_i^\dagger(s,w) \leq u, \Delta_i(s,w) = k \}$.
We propose a procedure that combines parametric bootstrap to take into account the variability associated to $\widehat \theta$ and perturbation-resampling methods, inspired by Sinnott and Cai (2016) \cite{sinnott2016inference}, to take into account the variability associated to $\widehat \Lambda_{k,0}(.)$. Indeed, parametric bootstrap does not apply to unspecified baseline risk functions.

This procedure, which also avoids hard computational cost, is realized as follows:\\[.2cm]
\fcolorbox{gray!30}{gray!30}{\parbox{\textwidth}{For each bootstrap sample $l=1,\ldots,L$, where $L$ is large enough;\\[.2cm]
\begin{minipage}{.98\textwidth}
\linespread{1.5}\selectfont
\begin{enumerate}
\item[1.] generate regression parameters from their asymptotic distribution $\widetilde\theta^{(l)} \sim \mathcal N(\widehat\theta, \widehat{V(\widehat\theta)})$;
\item[2.] deduce the empirical Bayes estimates $\widehat b_{.}^{(l)} = \mathbbm E(b_{.} | \mathcal Y_{.}(s), \mathcal X_{.}(s); \widetilde\theta^{(l)})$;
\item[3.] generate a perturbation $\nu_i^{(l)} \sim 4 \cdot \mathrm{Beta}({1}/{2}, {3}/{2})$ for each subject $i \in 1,\ldots, N^\dagger(s)$ of the learning sample;
\end{enumerate}
\end{minipage}}}\\
\fcolorbox{gray!30}{gray!30}{\parbox{\textwidth}{
\begin{minipage}{.98\textwidth}
\linespread{1.5}\selectfont
\begin{enumerate}
\item[4.] compute the Breslow's estimator $\widetilde \Lambda_{k,0}^{(l)}(u)$ from the perturbed hazard and the perturbed proportion of events at parameter values $\widetilde\theta^{(l)}$ and $\widehat b_{.}^{(l)}$:
\begin{align*}
& \widehat\Pi_k^{(0),(l)}(\widetilde\theta^{(l)},u)=\dfrac{1}{N^\dagger(s)} \sum\limits_{i=1}^{N^\dagger(s)} \nu_i^{(l)} \times \mathbbm 1\{ T_i^\dagger(s,w) \geq u\}  \exp\left\lbrace X^{E~\top}_{k,i} \widetilde\gamma_k^{(l)} + \widehat W_{k,i}(s | \widehat b_i^{(l)} ; \widetilde \beta^{(l)})^\top \widetilde\eta_k^{(l)} \right\rbrace  \\
& \bar J_k^{(l)}(u) = \dfrac{1}{N^\dagger(s)} \sum\limits_{i=1}^{N^\dagger(s)} \nu_i^{(l)}  \times \mathbbm 1\{ T_i^\dagger(s,w) \leq u, \Delta_i(s,w) = k \};
\end{align*}
\item[5.] deduce the predicted probability $\widetilde {\pi}_\star^{k,(l)}(s,w; \widetilde{\theta}^{(l)})$ as a function of $\widetilde\theta^{(l)}$, $\widehat b_\star^{(l)}$ and $\{(\widetilde \Lambda_{k,0}^{(l)}(u), s<u \leq s+w);k=1,\ldots,K\}$. See details in Section 1.4 of the Supplementary Material.
\end{enumerate}
Compute the 95\% confidence interval from the 2.5th and 97.5th percentiles of $\{ \widetilde{\pi}_\star^{k,(l)}(s,w; \widetilde{\theta}^{(l)}); l=1,\ldots,L \}$.
\end{minipage}}}\\

With the naive approach, the predicted conditional cumulative probability of cause $k$ is
\vspace{-.1cm}
\begin{equation}
\widehat\pi_\star^k(s,w ; \widehat\theta) = 
\Pr \, ( s < T_\star \leq s+w, \delta_\star = k | T_\star > s, \{X^E_{k,\star};k=1,\ldots,K\}, Y_\star(t_{\star n_\star(s)}); \widehat\theta )
\vspace{-.1cm}
\end{equation}
with $\widehat\theta=( \widehat\gamma^\top, \widehat\eta^\top )^\top$. \\
The same technique combining parametric bootstrap and perturbation-resampling can be used to obtain 95\% confidence intervals.

\subsection{Landmark model based on pseudo-observations}
\label{s:s:other}

Cause-specific hazard models rely on the PH assumption and require the computation of integrals over time in the individual cumulative probabilities. To avoid these issues, some authors have focused on the direct modelling of the individual cumulative probabilities, as for example the pseudo-value approach \citep{andersen2010pseudo}. As the latter does not require the PH assumption, the considered information is $\mathcal I = \{( T_i^\dagger, \Delta_i, \mathcal Y_i(s), \mathcal X_i(s) ) ; i=1,\ldots,N^\dagger(s) \}$.

\subsubsection{Model formulation}
\label{p:pv_model}
$~$\\
For subjects at risk at time $s$, we are interested in the expectation of
$\mu_{i}^k(s,w) = \mathbbm{1}(T_i \leq s+w, \delta_i = k)$.
In presence of censoring, this quantity is not always observable. Thus the idea is to define the dynamic jackknife pseudo-observation \citep{nicolaie2013dynamic} of the non-parametric estimator of $\pi^k(s,w)$: $\widehat{\mu}^k_{i}(s,w) =  N^\dagger(s)\widehat{F}^k(s,w) - (N^\dagger(s) - 1)\widehat{F}_{(-i)}^{k}(s,w)$, where $N^\dagger(s)$ is the number of subjects at risk at $s$ and $\widehat{F}^k(s,w)$ is the Aalen-Johansen estimate of $\pi^k(s,w)$.

To include the dynamic information on the marker until $s$, the same two-stage approach as defined in Section \ref{s:s:lm} can be used to deduce $\widehat W_{k,i}(s | \widehat b_i ; \widehat\beta)$ in those still at risk in $s$. The pseudo-observation and the prognostic factors are then linked through a generalized linear model with link function $g$:
\begin{align*}
g \left[ \mathbbm E\lbrace \widehat\mu^k_{i}(s,w) | T_i^\dagger > s \rbrace \right] = \gamma_{0,k} + X_{k,i}^{E~\top} \gamma_{1,k} + \widehat W_{k,i}(s | \widehat b_i ; \widehat\beta)^\top \eta_k.
\end{align*}
The model is thus estimated using generalized estimating equations (GEE).

\subsubsection{Cumulative probability estimator}
\label{p:pv_ci}

$~$\\
The predicted conditional cumulative probability can be directly expressed as
\begin{align}
\label{e:pv_est}
\widehat\pi_\star^k(s,w ; \widehat\theta) &= \Pr \, ( s < T_\star \leq s+w, \delta_\star = k | T_\star > s, \mathcal X_\star(s), \widehat{b}_\star ; \widehat\theta),
\end{align}
with $\widehat b_\star = \mathbbm E(b_\star | \mathcal Y_\star(s), \mathcal X_\star(s); \widehat \theta)$ and $\widehat\theta = ( \widehat\beta^\top, \widehat\sigma^{2}, \widehat\gamma_{0,k}, \widehat\gamma_{1,k}^\top, \widehat\eta_k^\top,$ $\textrm{vec}(\widehat D)^\top)^\top$ the vector of estimated parameters (with associated estimated variance matrix $\widehat{V(\widehat\theta)}$).
When considering the cloglog link function ($\textrm{cloglog}(x)=\log\{-\log(1-x)\}$) for instance, the predicted probability is $\widehat{\pi}^k_\star(s,w ; \widehat\theta) = 1 - \exp\Big[ -\exp\{ \widehat\gamma_{0,k} +$ $X_{k,i}^{E~\top} \widehat\gamma_{1,k} + \widehat W_{k,i}(s | \widehat b_i ; \widehat\beta)^\top \widehat\eta_k \}\Big]$.

The 95\% confidence intervals of \eqref{e:pv_est} may be calculated using parametric bootstrap:\\[.2cm]
\fcolorbox{gray!30}{gray!30}{\parbox{\textwidth}{Consider a large $L$; for each $l=1,\ldots,L$,\\[.2cm]
\begin{minipage}{.98\textwidth}
\linespread{1.5}\selectfont
\begin{enumerate}
\item[1.] generate parameters from their asymptotic distribution $\widetilde\theta^{(l)} \sim \mathcal N(\widehat\theta, \widehat{V(\widehat\theta)})$;
\item[2.] deduce the empirical Bayes estimates $\widehat b_\star^{(l)} = \mathbbm E(b_\star | \mathcal Y_\star(s), \mathcal X_\star(s); \widetilde\theta^{(l)})$;
\item[3.] compute the predicted probability $\widetilde {\pi}_\star^{k,(l)}(s,w; \widetilde{\theta}^{(l)})$ defined in \eqref{e:pv_est} for parameter values $\widetilde{\theta}^{(l)}$ and corresponding $\widehat b_\star^{(l)}$.
\end{enumerate}
Compute the 95\% confidence interval from the 2.5th and 97.5th percentiles of $\{ \widetilde{\pi}_\star^{k,(l)}(s,w; \widetilde{\theta}^{(l)}); l=1,\ldots,L \}$.
\end{minipage}}}\\

\subsection{Implementation}

The estimation of prediction models and the computation of derived estimators were done in R using standard packages and extensions coded by the authors, with the \texttt{JM} package for joint models, the \texttt{survival} package for the landmark cause-specific proportional hazards models and the \texttt{pseudo} and \texttt{geepack} packages for landmark models based on pseudo-values.
Examples of codes used for the simulations can be found in Supplementary Material, Section 5, and detailed examples can be found at \url{https://github.com/LoicFerrer/} for practical use.

\section{Illustrative example in Prostate Cancer progression}
\label{s:data}

We illustrate the computation of individual dynamic predictions in the context of localized Prostate Cancer treated by external beam radiotherapy. The objective was to predict the risk of progression after treatment from the prognosis factors collected at diagnosis (e.g., tumor stage) and the history of the Prostate Specific Antigen (PSA), the central marker in Prostate Cancer, repeatedly collected after treatment. We used the same data as analyzed in Ferrer et al. (2016) \cite{ferrer2016joint} from the multi-center clinical trial RTOG 9406 (USA) \cite{michalski2005toxicity} and the cohort of the British Columbia Cancer Agency in Vancouver (Canada) \cite{pickles2003evaluation}. 
Specifically, after the end of the radiotherapy, repeated PSA measurements were collected until the occurrence of a disease recurrence (local/distant recurrence, initiation of hormonal therapy or death due to prostate cancer) or death due to an other cause. After radiotherapy, post-treatment PSA trajectory is mostly biphasic with a short term drop associated with the treatment efficacy and possibly followed by a linear increase, associated with higher risk of recurrence. Some authors showed in this context that including the post-treatment PSA dynamics in dynamic prediction tools of overall disease recurrence highly improved the prediction accuracy compared to tools using only diagnosis information \citep{proust2009development, taylor2013real}.

All the models considered a biphasic trajectory of PSA and competing risks of recurrence and of death before recurrence. They were adjusted on the cohort (\texttt{Cohort}), age (\texttt{Age}), tumor stage (\texttt{Tstage}), Gleason score (\texttt{Gleason}, which measures aggressiveness of the cancer) and pretreatment log PSA level (\texttt{Initial\_logPSA}). The effect of the PSA dynamics on the specific hazards of recurrence and death was summarized by the extrapolated level and slope of the PSA in the joint models and two-stage landmark models.

We excluded two patients from the estimation sample and computed for them cumulative probabilities of having a prostate cancer recurrence in an horizon of 1.5 or 3 years from two landmark times. The trajectories of PSA and estimated dynamic predictions are displayed in Figure~\ref{f:illustrative_example} (Figure 1 of the Supplementary Material displays the predictions for death without recurrence). 

After one year and a half (1.3 years exactly for patient A), the two patients have roughly the same PSA trajectory. Yet, their predicted probabilities of recurrence substantially differ: patient B has a higher probability to progress than patient A; this is explained by worse prognostic factors at the end of treatment (higher tumour stage, Gleason score and initial PSA level). 

After one additional year, PSA trajectories strictly differ between the two patients: PSA has kept going down for patient B while PSA has sharply increased for patient A. These results in an updated predicted cumulative probabilities of recurrence systematically lower than $0.11$ for patient B and systematically higher than $0.70$ for patient A. In reality, patient A was diagnosed with a local recurrence 3 years after the end of radiotherapy while patient B was censored without any event after 13 years of follow-up.

Predictions from all the models are roughly in accordance. However, we observe that point estimates and width of confidence intervals substantially vary between models, especially when predictions are high. 

\section{Simulation studies}
\label{s:simu}

Two simulation studies were performed, one for the validation of the estimators (Section~\ref{s:s:simu1}), and a second for their comparison and the assessment of their robustness to misspecification (Section~\ref{s:s:simu2}). Both simulation studies relied on the Prostate Cancer example of Section~\ref{s:data} to generate realistic data and on the same following design.

$R=500$ learning samples of $N=1000$ subjects as well as one validation sample of $N^\textrm{new}(0) = 500$ subjects were generated from a joint model with parameter values $\theta_0$ \citep{ferrer2016joint}. The models detailed in Section \ref{s:prediction_models} were estimated on each learning sample $r$ ($r=1,...,R$) and the derived estimators of cumulative probability were computed for a given horizon $w$ on the $N^\textrm{new}(s)$ subjects ($\star = 1,\ldots,N^\textrm{new}(s)$) of the validation sample who did not experience any event before landmark time $s$. 
For each replicate $r$, we then compared the true generated cumulative probability $\pi_\star^k(s,w; \theta_0) = \int\limits_{\R^q} \pi_\star^k(s,w|b_\star;\theta_0) \, f(b_\star|T_\star > s, \mathcal Y_\star(s), \mathcal X_\star(s); \theta_0) \, \mathrm db_\star$ with the estimators  $\widehat\pi_{\star,r}^k(s,w; \widehat\theta)$.

\subsection{Simulation study \RN{1}: Validation of the estimators $\widehat\pi^{k}_\star(s,w;\widehat\theta)$}
\label{s:s:simu1}

We validated the estimators by checking the distributions over the individuals of estimated relative bias and coverage rates for $\pi_{\star}^k(s,w; \theta_0)$. We also investigated their efficiency using the mean over the replicates of the confidence interval widths.

\subsubsection{Model specification}
$ ~$ \\
For each subject of the learning or validation sample, data were generated according to a joint model with linear marker trajectory, true current level and true current slope of the marker as association structure $W_{k,.}(t|b_{.},\beta)$, and two causes of event (Recurrence ; Death). The model formulation is given in Supplementary Material, Section 3.1. The coefficients and covariate distributions used in the simulations correspond to those obtained on the motivating data.

\subsubsection{Results}
$ ~$ \\
Due to the duration of the procedures, the simulations were run for two landmark times $s=1$ and $s=5$, one horizon time $w=3$ and 200 subjects randomly selected from the validation sample. $R=499$ and $R=486$ replicates were considered for $s=1$ and $s=5$ respectively, due to convergence problems in the landmark model estimation.

Figures~\ref{f:bias} and~\ref{f:CR} depict respectively the distribution over the subjects of the relative bias of the estimator and the coverage rates of its 95\% confidence interval both for the joint and two-stage landmark CS PH models for landmark times $s=1$ and $s=5$ and one horizon time $w=3$.
The box plots highlight the correct estimation of $\pi_\star^{k}(s,w;\theta_0)$, except for the conditional expression from the joint model in the earlier landmark times ($s=1$). This confirms that considering the modes of the posterior distributions of the random effects (see Supplementary Material, Section 1.2) is valid only when there is enough longitudinal information.
The coverage rates which are very close to 0.95 validate the proposed 95\% confidence interval computations for both approaches. 
Finally the comparison of the widths of the 95\% confidence intervals according to the joint and two-stage landmark CS PH models (Figure \ref{f:CI_width}) confirms that the joint model estimator is much more efficient than the landmark CS PH estimator. 

\subsection{Simulation study \RN{2}: Robustness to models hypotheses}
\label{s:s:simu2}

The second simulation study aimed to compare the robustness of the approaches to model misspecification. We relied for this on predictive accuracy on the validation sample with both the Area Under the ROC curve (AUC) and the Mean Squared Error of Prediction (MSEP) popularized through the Brier Score (BS). Since we were in a simulation study, rather than using the event indicator and computing the BS, we directly used the true generated individual prediction and computed the MSEP:
 $\textrm{MSEP}^k_r(s,w) = \dfrac{1}{N^\textrm{new}(s)} \times \sum\limits_{\star=1}^{N^\textrm{new}(s)} \left( \pi_{\star}^k(s,w;\theta_0) - \widehat\pi_{\star,r}^k(s,w;\widehat\theta) \right)^2$. 
For the AUC, we applied the definition adapted to the competing risks setting \citep{blanche2015quantifying}: $\textrm{AUC}^k_r(s,w) = \Pr \, ( \widehat\pi_{i,r}^k(s,w;\widehat\theta) > \widehat\pi_{j,r}^k(s,w;\widehat\theta) \, | \, \Delta^k_i(s,w) = 1, T_i > s, \Delta^k_j(s,w)=0, T_j > s)$,
where $\Delta_i^k(s,w) = \mathbbm 1\{s<T_i\leq s+w, \delta_i=k\}$ with $\delta_i=k$ the cause of event ; $i$ and $j$ are here two subjects for prediction (see Supplementary Material 2.2 for details on the AUC formulation). Note that both AUC and MSEP estimators were intrinsically model free since we were in a simulation study and did not have to deal with censoring. 

We considered four scenarios: (1) correct specification of the joint model, (2) misspecification of the dependence function, (3) violation of the proportional hazards assumption, and (4) misspecification of the longitudinal trajectory of the marker. The distribution of the covariates and the coefficients used for the generation data in the four cases can be found in Supplementary Material, Section 3. Under each scenario, for the sake of scale and readability of the results, we chose to display boxplots of the differences in the predictive accuracy measures over the $R$ replicates to compare prediction models two by two; mean absolute measures are indicated for the marginal estimator taken as the reference in the comparisons. We focused on the error or prediction with MSEP in the main manuscript (Figures \ref{f:case1} to \ref{f:case4}) as it assesses both calibration and discrimination abilities of the methods. Results on AUC (which only focuses on discrimination ability and as such neglects an important aspect of predictive accuracy \citep{blanche2015quantifying}) are provided in Supplementary Material (Figures 6 to 10).

\subsubsection{Case 1: Correct specification of the joint model}
$~$\\
For the well-specified case, data generation and specification of the joint and landmark models in the estimation and prediction steps were the same as in Section~\ref{s:s:simu1}.

Figure~\ref{f:case1} shows differences of MSEP for 8 pairs of landmark and horizon times ($s=1,3,5,8$ and $w=1.5,3$).
As expected, the joint model performed better than the landmark models for all the pairs $(s,w)$. Once again, the conditional estimator of the predicted probability in the joint model was much worse than its marginal alternative in the earliest landmark times, but gave similar performances from $s=5$.
When assessing discriminatory power with AUC (Figure 6 in Supplementary Material), the joint model still performed better than landmark models, especially the naive landmark model. However, no difference was highlighted between the conditional and marginal estimators suggesting a problem of calibration rather than discrimination for the conditional estimator in the earliest times. It can be noted that most of the convergence problems in the model estimations arose from insufficiently considered information in landmarking.

\subsubsection{Case 2: Misspecification of the dependence function}
$~$\\
To investigate a misspecification of the dependence structure, we used the same generated data as in case 1 but the prediction models neglected the slope of the marker in the estimation and prediction steps although marker slope had a strong impact on the risk of recurrence.

The distributions over the replicates of the differences of MSEP for all the selected pairs of landmark and horizon times are depicted in Figure~\ref{f:case2}. 
Estimators behave similarly as in case 1 relative to the JM marginal estimator. However, neglecting the slope in the dependence structure induced a large increase in the MSEP of the JM marginal estimator (given by $\overline{\textrm{ref}}$ in Figure~\ref{f:case1} and Figure~\ref{f:case2}); for instance, for $(s,w) = (1,3)$, the MSEP increases from 0.323 to 1.114. This underlines the great importance of correctly specifying the dependence function in these models. The examination of AUC differences (Figure 7 in Supplementary Material) led to the same conclusions, except that the AUC under case 1 was not systematically better than the one under case 2.

\subsubsection{Case 3: Violation of the proportional hazards assumption}
$~$\\
The robustness of the models to a violation of the proportional hazard assumption was checked by considering an interaction with $\log(1+t)$ for the survival parameters associated with the marker dynamics in the generation model (see Section 3.2 of the Supplementary Material for the model specification).
For all the prediction models, the estimation and prediction steps did not consider this interaction with $\log(1+t)$.

Boxplots of the differences of MSEP over the replicates are depicted in Figure~\ref{f:case3}.
Even under this strong violation of the PH assumption, the performances of the two-stage landmark and joint models remained comparable. Indeed, although the landmark model permitted to obtain estimated parameters closer to the generated one, their variances were very large because of the much reduced information used in these models (see Figure~2 in Supplementary Material for an illustration of the estimators behavior).
One can also note that the pseudo-value approach was not better than the models based on proportional hazards. 
Again, the same conclusions were drawn from the differences in AUC (Figure 8 in Supplementary Material). 

\subsubsection{Case 4: Misspecification of the longitudinal trajectory of the marker}
$~$ \\
The last case explored the performances of the prediction models when the longitudinal trend of the marker was misspecified.
Data were generated using a joint model with a biphasic shape of the marker as generally observed for PSA data, whereas a linear trajectory over time for the marker was considered for the estimation of the predicted probabilities of event using joint and two-stage landmark models (see Section 3.3 of the Supplementary Material for the model specification).
The degree of misspecification of the longitudinal marker trend (shown in Figure~3 of the Supplementary Material) was chosen to be severe to clearly show the impact of such misspecification.

Figure~\ref{f:case4} displays the boxplots of differences in MSEP for the 8 pairs $(s,w)$. The landmark models performed much better than the joint models for landmark times $s=1,3,5$; at landmark time $s=8$, performances of joint and landmark models became roughly similar. Such result was expected. The joint model incorrectly assumed a linear trajectory for the marker on the whole follow-up while the landmark model, by considering only the longitudinal information collected until $s$, assumed a linear trajectory only until $s$ which was more realistic at earliest landmark times even if still far from being well specified.
The same conclusions were drawn from the examination of the AUC differences (Figure 9 in Supplementary Material), except that even at the latest landmark time (8 years), the landmark models remained much more discriminatory than the joint model.


To explore whether such differences were due to the severe misspecification of our example, we considered a second longitudinal marker trend (see Section 2.1 of the Supplementary Material for model specification). This supplementary case considered a small degree of misspecification of the longitudinal marker by considering some slight fluctuations with splines in the generation model compared to the well-specified case 1. Although only slightly misspecified, the superiority of joint model over landmark approaches previously found in case 1 almost completely disappeared. This confirmed the high sensitivity to any kind of misspecification of the marker trajectory in the joint model.

\section{Discussion}
\label{s:discussion}

With the development of personalized medicine, it is important to provide valid and powerful tools to clinicians for the computation of individual probabilities of specific events such as landmark conditional cumulative probabilities. These predictions are expected to be used in clinical practice, notably to adapt individual strategies of treatment or to plan the patient-specific optimal screening time in clinical trials.\\

Several authors \citep{rizopoulos2011dynamic, maziarz2017longitudinal} already proposed estimators of the individual conditional cumulative probability $\pi_\star^k(s,w)$, but none was formally validated.
Our first objective was thus to precisely define the quantity of interest $\pi_\star^k(s,w)$ and provide estimators (along with 95\% confidence interval) both for the landmarking and the joint modelling approaches and properly validate them by comparing generated and estimated expressions of $\pi_\star^k(s,w)$. The generated $\pi_\star^k(s,w)$ was not obvious to compute as involving an integral over the latent structure shared by the longitudinal and survival processes, the data being generated from a joint model. Note that such quantity is not always correctly defined (e.g., \citep{barrett2017dynamic, rizopoulos2017dynpred}). 
The marginal estimator from the joint model obtained very good performance in general whereas the conditional estimator from the joint model was found to necessitate substantial individual longitudinal information collected until the prediction time to be accurate.

Quantification of the uncertainty around individual predictions is essential for the decision making in clinical practice.
In the landmark approach no solution was ever proposed, and a vagueness prevailed in the joint modelling literature with some definitions conditional on the marker's observations \citep{proust2009development, taylor2013real, maziarz2017longitudinal}, others taking into account the measurement error in the marker's observations either along with the population parameter uncertainty \citep{yu2008individual,rizopoulos2011dynamic} or without \citep{desmee2017nonlinear}. Since we defined a quantity of interest conditional on the marker's observations, our definitions of uncertainty only took into account the variability due to parameter estimates. We proposed corresponding Monte Carlo methods to compute the confidence intervals of the predictions in a unified manner for joint and landmark models and we showed that they correctly assessed the uncertainty around the individual predictions using a simulation study.
Compared to the joint model, the estimator based on the two-stage landmark cause-specific proportional hazards model confirmed its expected poor efficiency, with wide confidence intervals when only a few subjects experienced the event in the prediction window.

Our second objective was to properly compare the landmark models and the joint model through several cases of well- and mis-specification. Indeed, a series of papers showed comparisons of prediction models in dynamic predictions \citep{goldstein2017comparison, huang2016two,sweeting2017use} but none of them evaluated their robustness to misspecifications although most proposed methods were parametric. To our knowledge, only one contribution explored the problem of misspecification of prediction models (only longitudinal trajectory and functional dependency) very recently in a small simulation study and concluded to the superiority of joint models approach over landmark models \citep{rizopoulos2017dynpred}.

In our extensive simulation study, we found that in the case of correctly specified model, the joint model performed better than the landmark models, as expected. In the case of misspecification of the dependence structure between the longitudinal process and the survival process, the difference of performances between approaches did not change as also concluded by Rizopoulos et al. (2017) \cite{rizopoulos2017dynpred}. But more importantly, the performances of misspecified models (both from joint modelling and landmarking approaches were much worse in terms of prediction error.
Regarding  the  PH  assumption, the two landmark models we proposed better dealt with this assumption than the joint model: dynamic pseudo-values did not require the PH assumption at all, and our cause-specific hazards landmark models limited the PH assumption to the prediction window with an administrative censoring at the end of it. Yet, the impact of PH assumption violation on the estimators derived from the joint model remained limited, suggesting that the violation of the PH assumption should be extreme to entail a tangible impact on the estimated cumulative probabilities in the joint model.
Finally, we showed that the correct specification of the marker trajectory was essential to provide good predictions with joint models (and with landmark models to a much lesser extent). We demonstrated the major loss of performance of the joint model in a severe case of misspecification to illustrate the limit but we also found in Supplementary Material that even a slight misspecification of the trajectory (usually considered as acceptable) impacted the prediction error of the models, and eliminated for example the gain of using the joint model over the landmark model at shorter landmark times or when the horizon time increased. 

As usual in prediction model development, comparisons were made in terms of both Mean Square Error of Prediction which measures a trade-off between calibration and discrimination, and Area Under the ROC Curve which only targets discriminatory power. 
With the perspective in mind of providing quantified individual predictions, we chose to primarily rely on the prediction error even though most conclusions were also drawn from AUC examinations, yet sometimes to a lesser extent. This was probably explained by the lesser sensitivity of AUC \citep{pencina2008evaluating} and the possibly preserved discriminatory power in the presence of worse calibration.\\

To conclude with recommendations, we emphasize the need to carefully define the quantity of interest, its estimator and the type of associated uncertainty. The several cases of misspecification warned us on the necessity to precisely specify the dependence structure between the longitudinal marker dynamics and the risk of event, and not systematically consider only the current marker level. Finally, the specification of the longitudinal marker trend should be studied with extreme care, especially when using joint models. Researchers should be warned that the use of sophisticated methods such as the joint models may allow obtaining accurate and efficient estimators only when they are correctly specified. Otherwise, estimators might be off the mark. Landmark models seem less sensitive to the misspecification of the longitudinal marker trajectory but they are as sensitive as joint models regarding the misspecification of the dependence structure. In addition, they provide considerably less efficient estimators and may induce convergence problems, notably when the landmark time increases and thus the considered information becomes too poor.

\section{Supplementary Material}
An Appendix may be found in the source package of this article on arXiv. Detailed examples of the
code can be found at \url{https://github.com/LoicFerrer}
for practical use.

\section*{Acknowledgements}
The authors thank warmly Boris P. Hejblum for his valuable advice and assistance.
This work was supported by a joint grant from INSERM and R\'egion Aquitaine, a grant from the Institut OpenHealth and a grant from the R\'eseau Franco-N\'eerlandais.
Computer time for this study was provided by the computing facilities MCIA (M\'esocentre de Calcul Intensif Aquitain) of the Universit\'e de Bordeaux and of the Universit\'e de Pau et des Pays de l'Adour.

\label{LastPage}

\bibliographystyle{biom}
\bibliography{Paper}

\clearpage

\begin{figure}
\begin{center}
\captionsetup{width=.85\linewidth}
\begin{tabular}{c}
\subfloat[Patient A: $\texttt{Cohort} = \textrm{RTOG 9406}$, $\texttt{Age} = 72$, $\texttt{Tstage}=2$, $\texttt{Gleason}=6$ and $\texttt{Initial\_logPSA}=0.74$.]{
\begin{tikzpicture}
\node[inner sep=0pt] (russell) at (0,0)
    {\includegraphics[width=.95\textwidth]{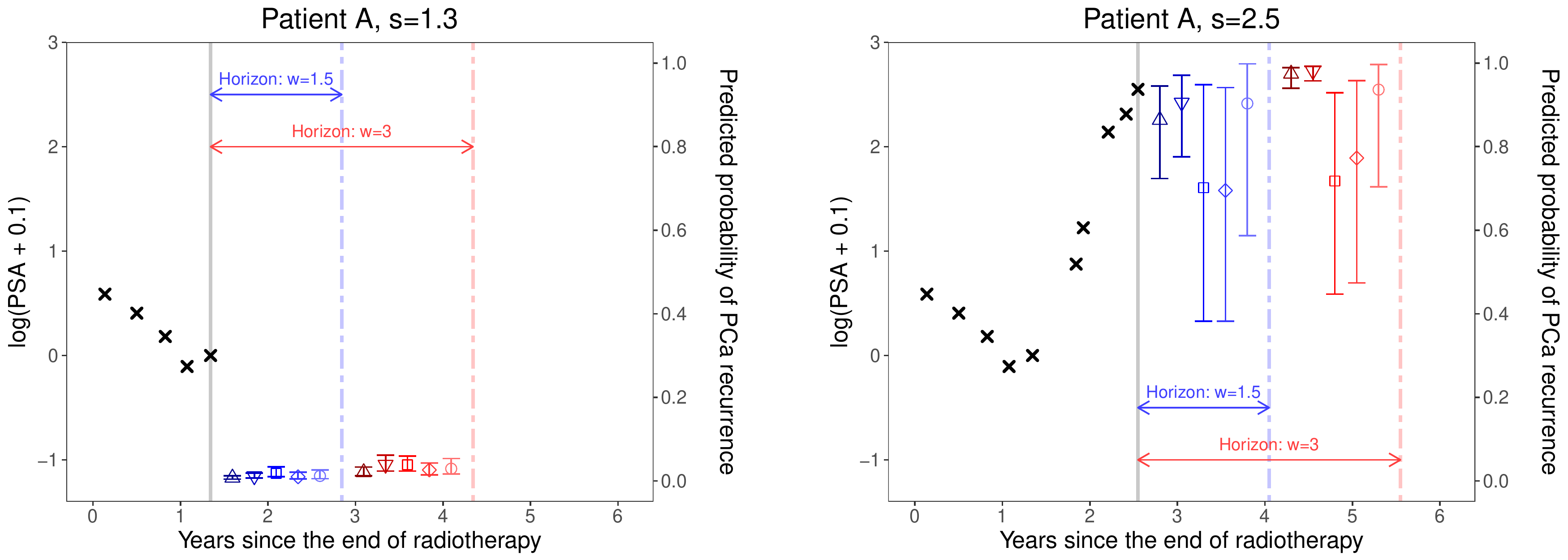}};
\node[inner sep=0pt] (whitehead) at (0,-3.0)
    {\includegraphics[width=.55\textwidth]{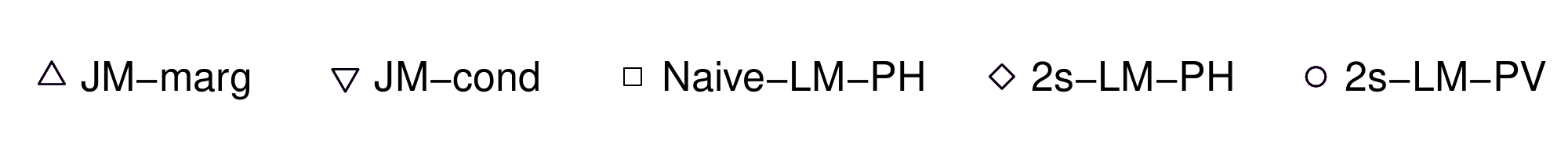}};
\end{tikzpicture}
\label{f:illustration1}}
 \\
\subfloat[Patient B: $\texttt{Cohort} = \textrm{RTOG 9406}$, $\texttt{Age} = 69$, $\texttt{Tstage}=3$, $\texttt{Gleason}=7$ and $\texttt{Initial\_logPSA}=2.07$.]{
\begin{tikzpicture}
\node[inner sep=0pt] (russell) at (0,0)
    {\includegraphics[width=.95\textwidth]{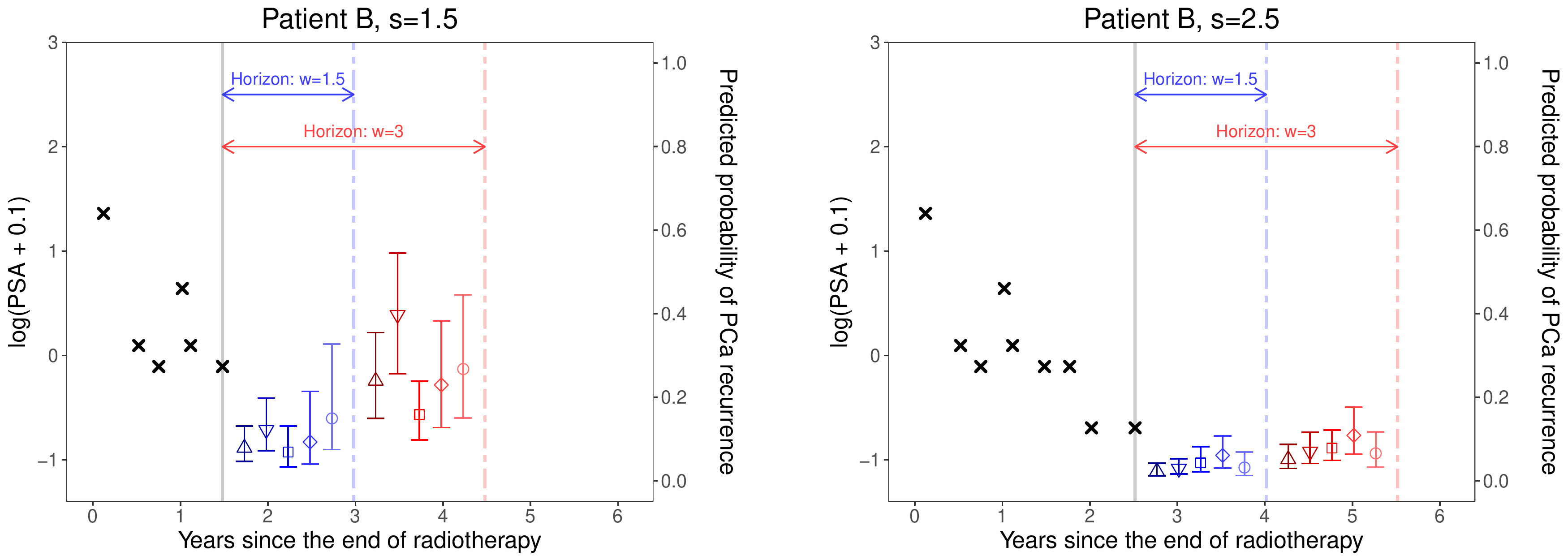}};
\node[inner sep=0pt] (whitehead) at (0,-3.0)
    {\includegraphics[width=.55\textwidth]{Legend_illustration}};
\end{tikzpicture}
\label{f:illustration2}}
\end{tabular}
\vspace{-.3cm}
\end{center}
\caption{Predicted individual cumulative probabilities of prostate cancer recurrence before death in two selected patients with a localized prostate cancer treated by radiotherapy (patient A for top panels and patient B for bottom panels). The predictions are computed from the log-levels of PSA collected until landmark time ($\times$) and baseline information using the marginal and conditional estimators from the joint model (\texttt{JM-marg} and \texttt{JM-cond}, respectively), the cause-specific landmark models with a naive or two-stage approach (\texttt{Naive-LM-PH} and \texttt{2s-LM-PH}, respectively) and the two-stage pseudo-value model (\texttt{2s-LM-PV}). For each subject, the probabilities are computed from two landmark times (plain grey lines at 1.3 or 1.5 for left panels and 2.5 for right panels and for two prediction windows $w=1.5$ and $w=3$ (dotted blue and red lines, respectively). The associated 95\% confidence intervals are obtained using 500 bootstrap samples.}
\label{f:illustrative_example}
\end{figure}

\begin{figure}[p]
\vspace{-.3cm}
\begin{center}
\captionsetup{width=.85\linewidth}
\begin{tabular}{c}
\subfloat[Evaluation of the estimators: distribution over the individuals $\star = 1,\ldots, N(s)$ of the relative bias (RB, in \%). The dashed line represents the 0.]{
\begin{tikzpicture}
\node[inner sep=0pt] (russell) at (0,0)
    {\includegraphics[width=.55\textwidth]{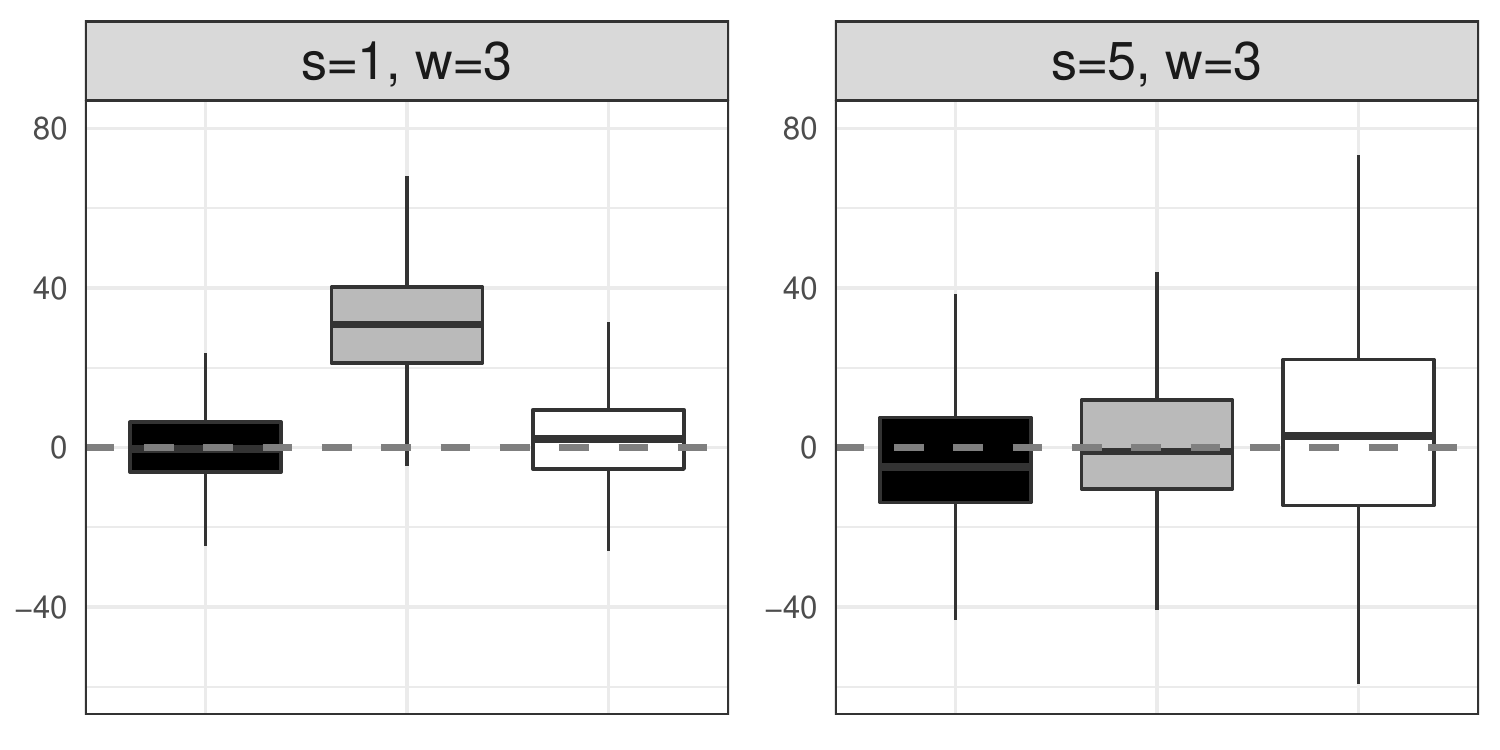}};
\node[inner sep=0pt] (whitehead) at (0,-2.4)
    {\includegraphics[width=.5\textwidth]{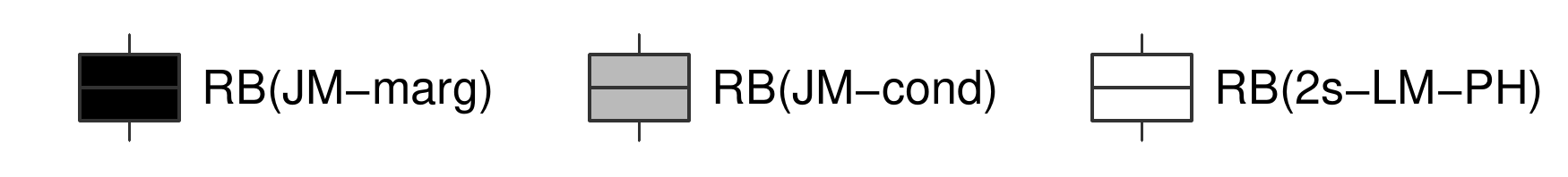}};
\end{tikzpicture}
\label{f:bias}} \\
\subfloat[Evaluation of the confidence intervals: distribution over the individuals $\star=1,\ldots,200$ of the coverage rates (CR) for the 95\% confidence intervals of $\pi_\star^{\textrm{Rec.}}(s,w; \theta)$. The dashed line represents the 0.95.]{
\begin{tikzpicture}
\node[inner sep=0pt] (russell) at (-0.8,0)
    {\includegraphics[width=.55\textwidth]{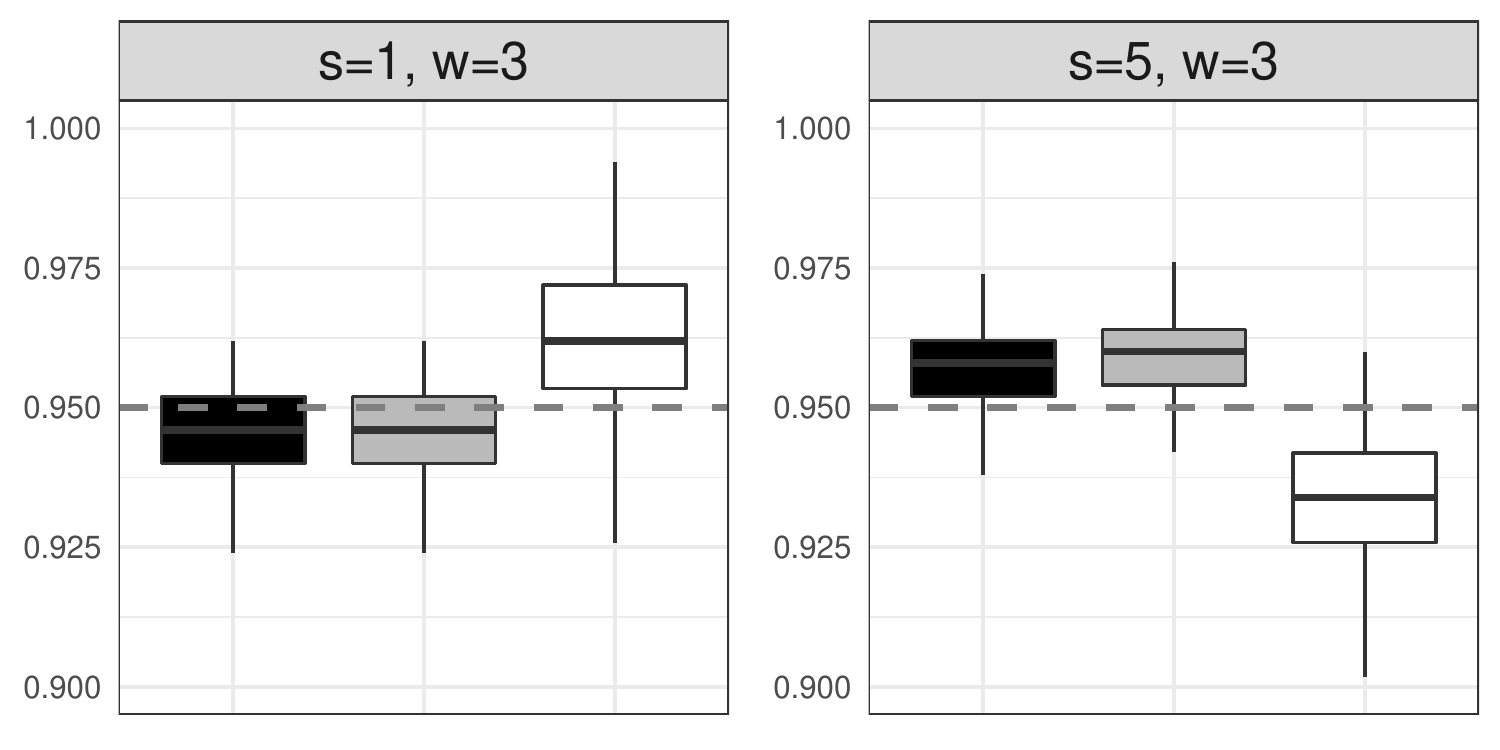}};
\node[inner sep=0pt] (whitehead) at (-0.6,-2.4)
    {\includegraphics[width=.5\textwidth]{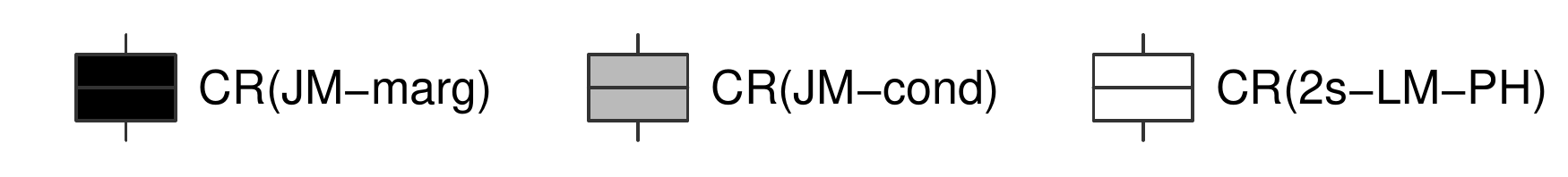}};
\end{tikzpicture}
\label{f:CR}} \\
\subfloat[Evaluation of the estimator efficiency: distribution over the individuals $\star=1,\ldots,200$ of the mean 95\% confidence interval widths ($|$CI$|$).]{
\begin{tikzpicture}
\node[inner sep=0pt] (russell) at (0,0)
    {\includegraphics[width=.55\textwidth]{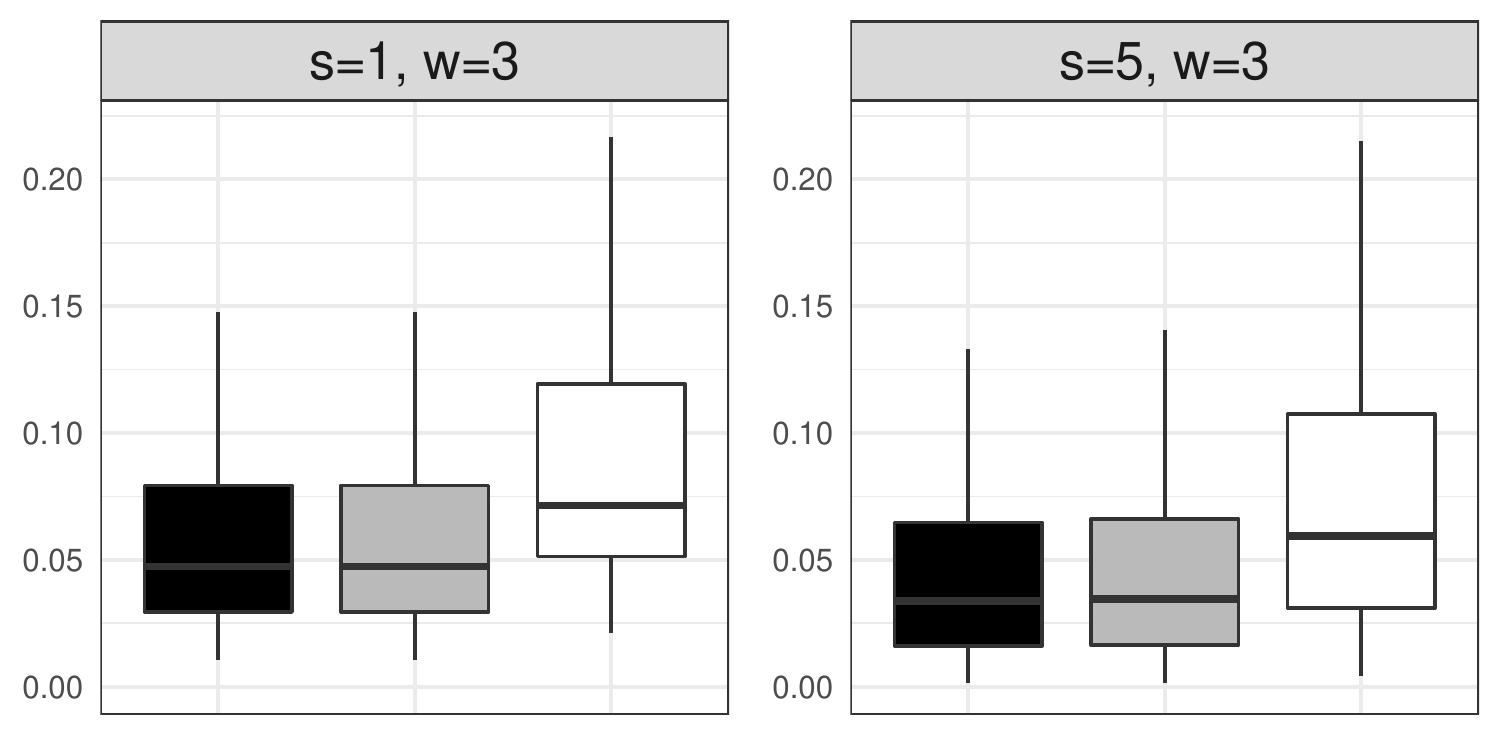}};
\node[inner sep=0pt] (whitehead) at (0,-2.4)
    {\includegraphics[width=.72\textwidth]{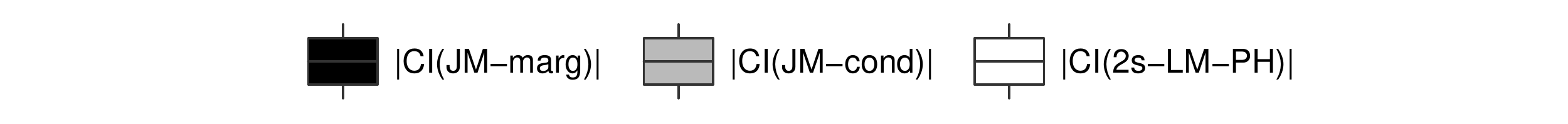}};
\end{tikzpicture}
\label{f:CI_width}}
\end{tabular}
\vspace{-.3cm}
\end{center}
\caption{Evaluation of the estimators in terms of relative bias (a), coverage rates (b) and confidence intervals widths (c). Considered are the marginal estimator and the conditional estimator from the joint model (denoted \texttt{JM-marg} and \texttt{JM-cond}, respectively) and the estimator based on the two-stage cause-specific landmark model (denoted \texttt{2s-LM-PH}).}
\end{figure}

\begin{figure}
\begin{center}
\vspace{-.5cm}
\begin{tikzpicture}
\node[inner sep=0pt] (russell) at (0,0)
    {\includegraphics[width=.75\textwidth]{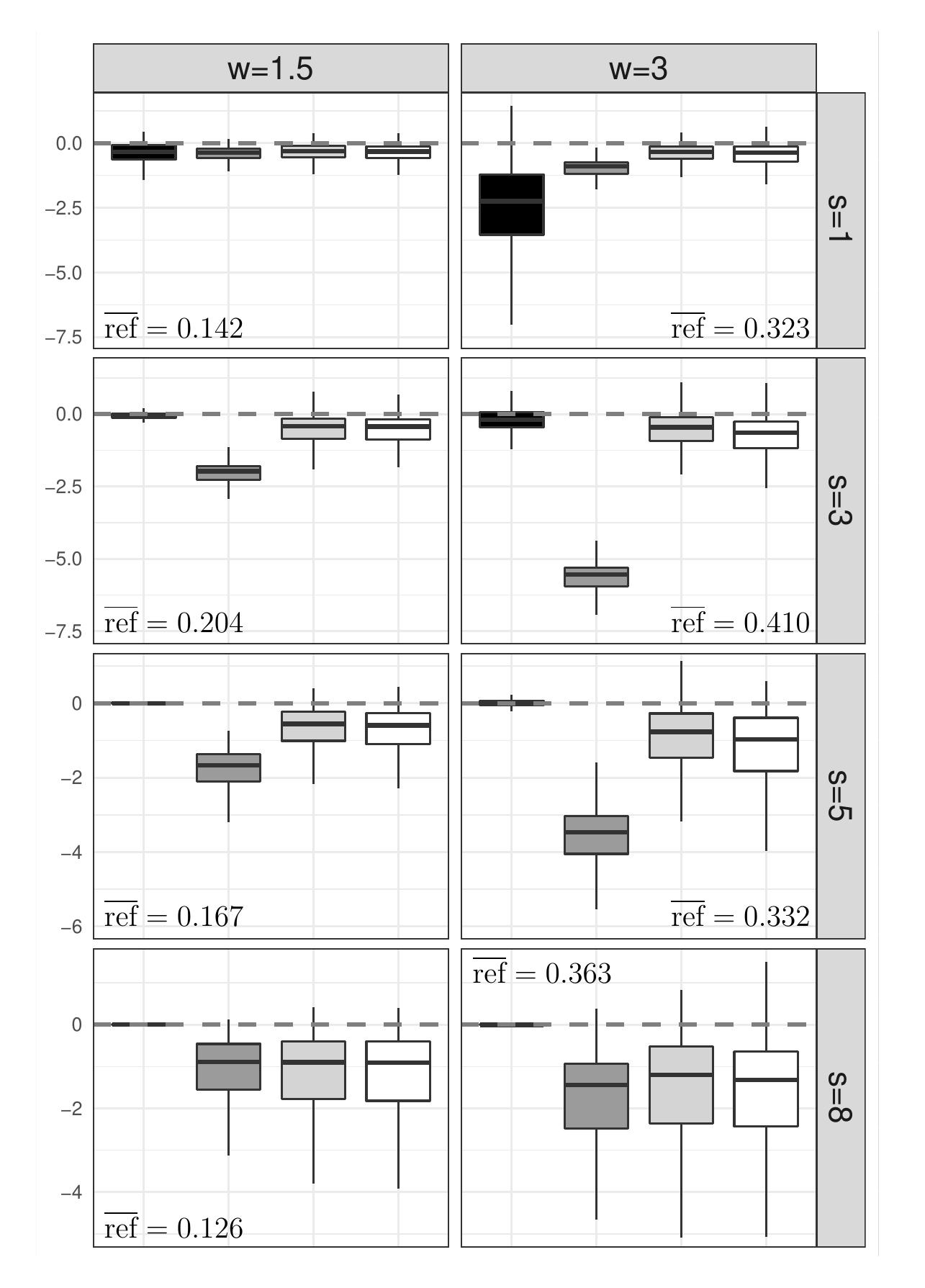}};
\node[inner sep=0pt] (whitehead) at (0,-8.4)
    {\includegraphics[width=.9\textwidth]{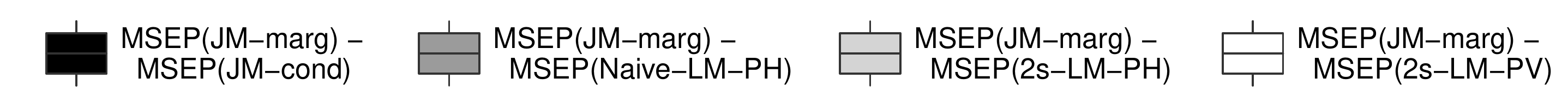}};
\end{tikzpicture}
\end{center}
\vspace{-.4cm}
\caption{
Boxplots of the differences ($\times 1000$) of Mean Square Error of Prediction (MSEP) between the marginal estimator from the joint model (denoted JM-marg) and alternatives in the case of correct specification of the joint model (case 1).
Considered are the conditional estimator from the joint model (JM-cond), the estimators from cause-specific landmark models using a two-stage or naive approach (2s-LM-PH and Naive-LM-PH, respectively) and the two-stage pseudo value model (2s-LM-PV). Only models that converged were considered.
The distributions are depicted over $R=499, 494, 486, 389$ replicates for 4 landmark times $s=1,3,5,8$ respectively, with 2 considered horizons $w=1.5$ and $w=3$.
$\overline{\textrm{ref}}$ denotes the mean MSEP ($\times 1000$) using the marginal estimator from the joint model for each $(s,w)$.
See the mean sample sizes and the mean number of recurrences occurred between $s$ and $s+w$ in Table~1 of the Supplementary Material.}
\label{f:case1}
\end{figure}
\begin{figure}
\begin{center}
\vspace{-.5cm}
\begin{tikzpicture}
\node[inner sep=0pt] (russell) at (0,0)
    {\includegraphics[width=.75\textwidth]{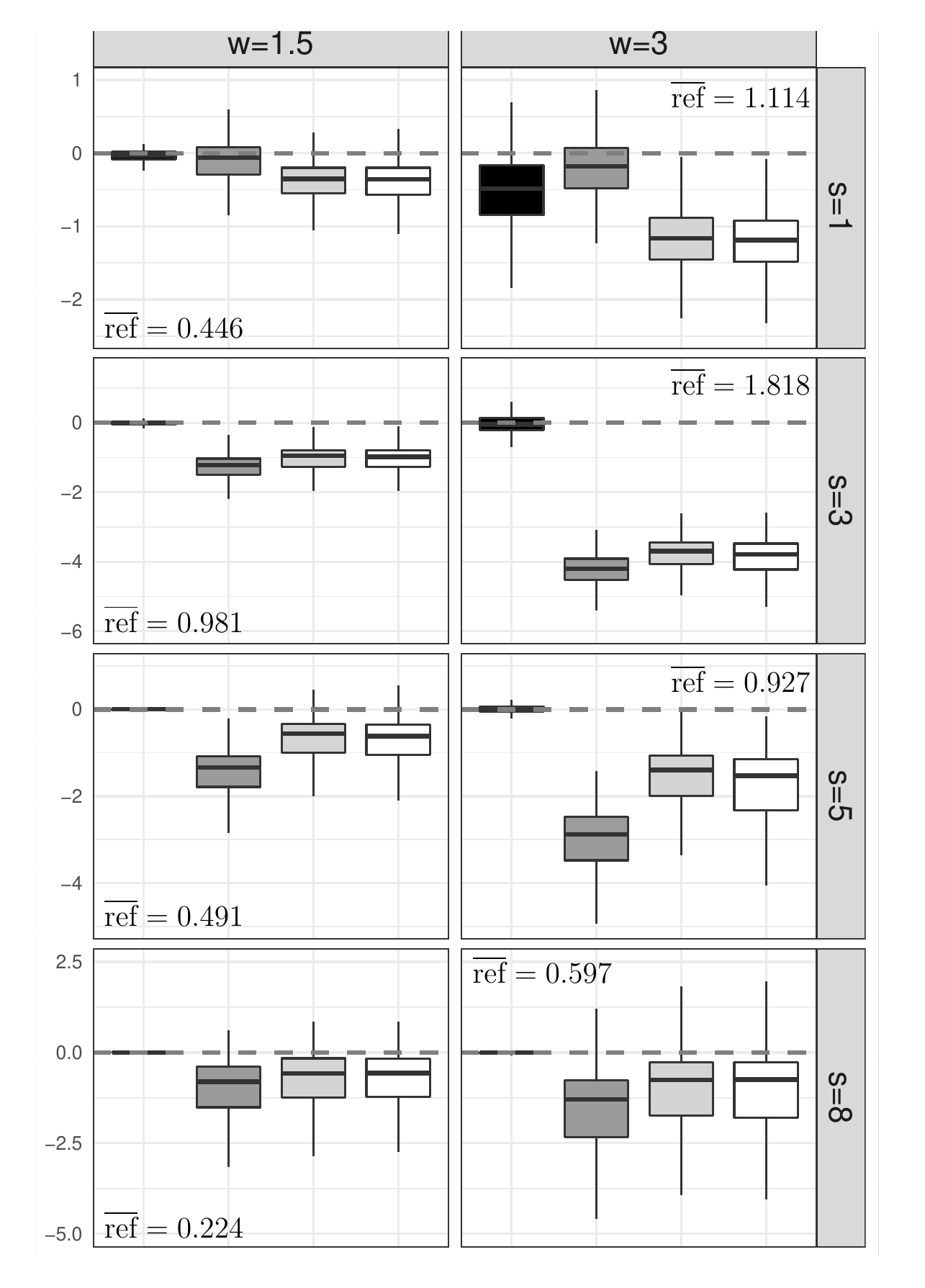}};
\node[inner sep=0pt] (whitehead) at (0,-8.4)
    {\includegraphics[width=.9\textwidth]{legend_test}};
\end{tikzpicture}
\end{center}
\vspace{-.4cm}
\caption{Boxplots of the differences ($\times 1000$) of Mean Square Error of Prediction (MSEP) between the marginal estimator from the joint model (denoted JM-marg) and alternatives in the case of misspecification of the dependence function (case 2).
Considered are the conditional estimator from the joint model (JM-cond), the estimators from cause-specific landmark models using a two-stage or naive approach (2s-LM-PH and Naive-LM-PH, respectively) and the two-stage pseudo value model (2s-LM-PV).
Only models that converged were considered.
The distributions are depicted over $R=499, 498, 497, 428$ replicates for 4 landmark times $s=1,3,5,8$ respectively, with 2 considered horizons $w=1.5$ and $w=3$.
$\overline{\textrm{ref}}$ denotes the mean MSEP ($\times 1000$) using the marginal estimator from the joint model for each $(s,w)$.
See the mean sample sizes and the mean number of recurrences occurred between $s$ and $s+w$ in Table~1 of the Supplementary Material.}
\label{f:case2}
\end{figure}
\begin{figure}
\begin{center}
\vspace{-.5cm}
\begin{tikzpicture}
\node[inner sep=0pt] (russell) at (0,0)
    {\includegraphics[width=.75\textwidth]{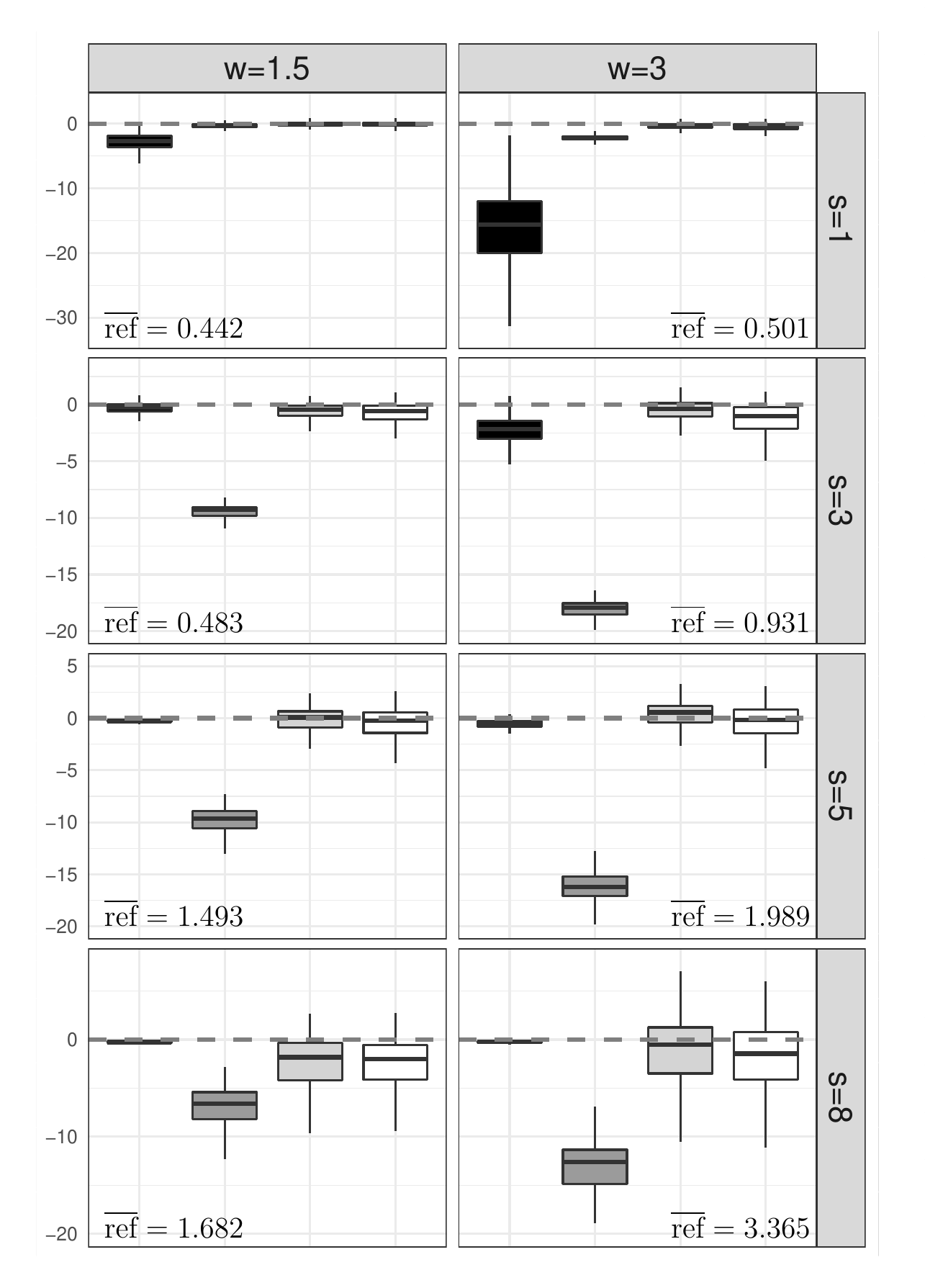}};
\node[inner sep=0pt] (whitehead) at (0,-8.4)
    {\includegraphics[width=.9\textwidth]{legend_test}};
\end{tikzpicture}
\end{center}
\vspace{-.4cm}
\caption{Boxplots of the differences ($\times 1000$) of Mean Square Error of Prediction (MSEP) between the marginal estimator from the joint model (denoted JM-marg) and alternatives in the case of substantial violation of the PH assumption (case 3).
Considered are the conditional estimator from the joint model (JM-cond), the estimators from cause-specific landmark models using a two-stage or naive approach (2s-LM-PH and Naive-LM-PH, respectively) and the two-stage pseudo value model (2s-LM-PV).
Only models that converged were considered.
The distributions are depicted over $R=485, 326, 294, 188$ replicates for 4 landmark times $s=1,3,5,8$ respectively, with 2 considered horizons $w=1.5$ and $w=3$.
$\overline{\textrm{ref}}$ denotes the mean MSEP ($\times 1000$) using the marginal estimator from the joint model for each $(s,w)$.
See the mean sample sizes and the mean number of recurrences occurred between $s$ and $s+w$ in Table~1 of the Supplementary Material.}
\label{f:case3}
\end{figure}
\begin{figure}
\begin{center}
\vspace{-.5cm}
\begin{tikzpicture}
\node[inner sep=0pt] (russell) at (0,0)
    {\includegraphics[width=.75\textwidth]{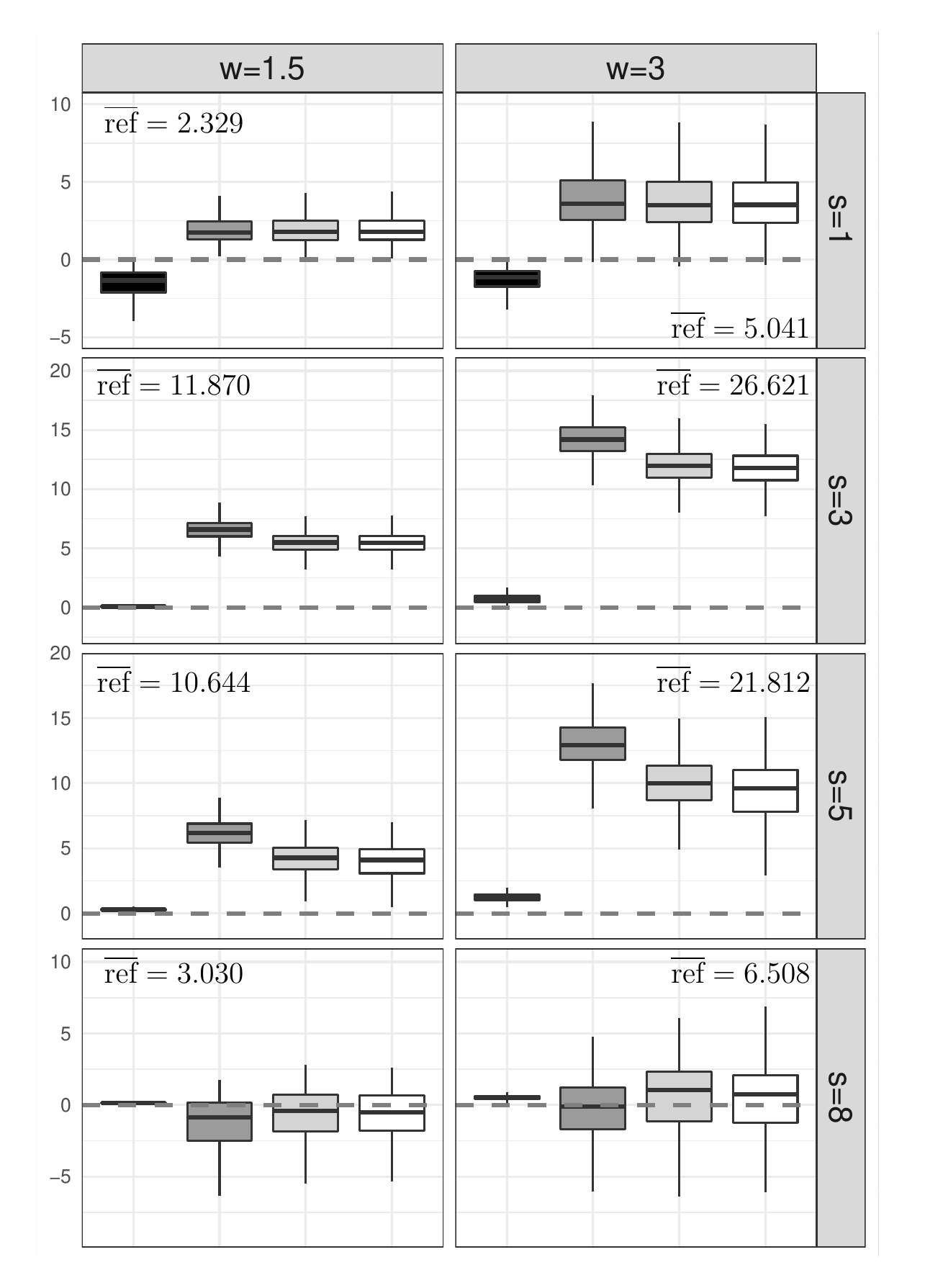}};
\node[inner sep=0pt] (whitehead) at (0,-8.4)
    {\includegraphics[width=.9\textwidth]{legend_test}};
\end{tikzpicture}
\end{center}
\vspace{-.4cm}
\caption{Boxplots of the differences ($\times 1000$) of Mean Square Error of Prediction (MSEP) between the marginal estimator from the joint model (denoted JM-marg) and alternatives in the case of substantial misspecification of the longitudinal marker trajectory (case 4).
Considered are the conditional estimator from the joint model (JM-cond), the estimators from cause-specific landmark models using a two-stage or naive approach (2s-LM-PH and Naive-LM-PH, respectively) and the two-stage pseudo value model (2s-LM-PV).
Only models that converged were considered.
The distributions are depicted over $R=500, 496, 476, 357$ replicates for 4 landmark times $s=1,3,5,8$ respectively, with 2 considered horizons $w=1.5$ and $w=3$.
$\overline{\textrm{ref}}$ denotes the mean MSEP ($\times 1000$) using the marginal estimator from the joint model for each $(s,w)$.
See the mean sample sizes and the mean number of recurrences occurred between $s$ and $s+w$ in Table~1 of the Supplementary Material.}
\label{f:case4}
\end{figure}

\end{document}